\documentclass[prb,
12pt,
superscriptaddress,showpacs,amsmath,amssymb]{revtex4}
\usepackage{amsfonts}
\usepackage{bm}
\usepackage{verbatim}

\usepackage{graphicx}

\begin{document}

\author{G.E.~Volovik}
\affiliation{Low Temperature Laboratory, Aalto University,  P.O. Box 15100, FI-00076 Aalto, Finland}
\affiliation{Landau Institute for Theoretical Physics, acad. Semyonov av., 1a, 142432,
Chernogolovka, Russia}

\title{Planck constants in the symmetry breaking quantum gravity}

\newcommand{\nh}{\slash\!\!\!h}

\date{\today}

\begin{abstract}
We consider the theory of quantum gravity  \cite{Akama1978,Diakonov2011}, in which gravity emerges as a result of the symmetry breaking transition in the quantum vacuum. The gravitational tetrads, which play the role of the order parameter in this transition, are represented  by the bilinear combinations  of the fermionic fields. In this quantum gravity scenario the interval $ds$ in the emergent general relativity is dimensionless.  Several other approaches to quantum gravity, including the model of superplastic vacuum  and $BF$-theories of gravity support  this suggestion. The important consequence of such metric dimension is that all the diffeomorphism invariant quantities are dimensionless for any dimension of spacetime \cite{Volovik2021,Volovik2021cont}. These include the action $S$, cosmological constant $\Lambda$, scalar curvature $R$, scalar field $\Phi$, wave function $\psi$, etc.  The composite fermion approach to quantum gravity suggests that the Planck constant $\hbar$ can be the parameter of the Minkowski metric \cite{Volovik2023a}. Here we extend this suggestion by introducing two Planck constants, bar $\hbar$ and slash $\slash\!\!\!h$, which are the parameters of the correspondingly time component and space component
of the Minkowski metric, $g^{\mu\nu}_{\rm Mink} = {\rm diag}(-\hbar^2,\nh^2,\nh^2,\nh^2)$.  The parameters bar $\hbar$ and slash $\nh$ are invariant only under $SO(3)$ transformations, and thus they are not diffeomorphism invariant.   As a result they have nonzero dimensions -- the dimension of time for $\hbar$ and dimension of length for $\nh$. Then according to the Weinberg criterion these parameters are not  fundamental, and may vary. In particular, they may depend on the Hubble parameter in expanding Universe.
They also change sign at the topological domain walls resulting from the symmetry breaking.
\end{abstract}

\maketitle

\newpage
\tableofcontents

\section{Introduction}

It is becoming clear that quantum gravity cannot be obtained by quantization of the classical gravity. Gravity can arise as emergent low-energy phenomenon, which comes from underlying quantum fields of the quantum vacuum. The typical example is provided by condensed matter, where the effective gravity emerges in the topological Weyl and Dirac materials:  semimetals, superfluids and superconductors. Gravitational tetrads emerge there in the vicinity of the conical points in the spectrum of fermionic quasiparticles \cite{Volovik2003}, see also recent papers \cite{Sabsovich2022,Chernyavsky2022}. Another condensed matter example of effective gravity is provided by the B-phase of superfluid $^3$He, where vielbein emerge as bilinear combinations of the fermionic fields \cite{Volovik1990}. Similar mechanism of the formation of the composite tetrads in the low energy physics has been suggested in the relativistic quantum field theories \cite{Akama1978,Diakonov2011,VladimirovDiakonov2012,VladimirovDiakonov2014,ObukhovHehl2012}. 
The emergent tetrads give rise to the effective metric (the four fermions object), to the interval, and finally to the effective action for the gravitational field. The important consequence of this mechanism is that all the diffeomorphism invariant physical quantities are dimensionless. Here we discuss one more consequence of such dimensionless physics, which is related to the Planck constant, and actually to two Planck constants, bar $\hbar$ and slash 
$\nh$. Both are the elements of the metric and tetrads in Minkowski vacuum. The bar $\hbar$ is the time component of the tetrad and has dimension of time, while the slash 
$\nh$ enters the space components of tetrads and has dimension of length.

\section{Composite tetrads from relative symmetry breaking}

The gravitational tetrads may appear as composite objects made of the more fundamental fields, the quantum fermionic fields \cite{Akama1978,Diakonov2011,VladimirovDiakonov2012,VladimirovDiakonov2014,ObukhovHehl2012,Maiezza2022}: 
\begin{equation}
 \hat E^a_\mu = \frac{1}{2}\left( \Psi^\dagger \gamma^a\partial_\mu  \Psi -  \Psi^\dagger\overleftarrow{\partial_\mu}  \gamma^a\Psi\right) \,.
\label{TetradsFermionsOperators}
\end{equation}
The original action does not depend on tetrads and metric and is described solely in terms of differential forms:
\begin{equation}
S=\frac{1}{24}e^{\alpha\beta\mu\nu} e_{abcd} \int d^4x  \, \hat E^a_\alpha   \hat E^b_\beta \hat E^c_\mu \hat E^d_\nu \,.
\label{OriginalActions}
\end{equation}
This action, which is the operator analog of the cosmological term, has high symmetry. It is symmetric under coordinate transformations $x^\mu \rightarrow \tilde x^\mu(x)$, and thus is also scale invariant. In addition, the action is symmetric under spin rotations, or under the corresponding gauge transformations when the spin connection is added to the gradients. The action may also contain the operator analog of the Einstein–Hilbert–Cartan term\cite{Diakonov2011}, $e^{\alpha\beta\mu\nu} e_{abcd} \int d^4x  \,\hat E^a_\alpha   \hat E^b_\beta F^{cd}_{\mu\nu}$, where $F^{cd}_{\mu\nu}$ is the Cartan curvature 2-form. Also the 4-form field can be included, $e^{\alpha\beta\mu\nu} e_{abcd} \int d^4x \,  F^{abcd}_{\alpha\beta\mu\nu}$, which is also related to the problem of the vacuum energy and cosmological constant \cite{3formHawking,KlinkhamerVolovik2008b,KlinkhamerVolovik2016,Nitta2018,Kaloper2022}.

The tetrads $e^a_\mu$ appear as the vacuum expectation values of the bilinear fermionic 1-form $\hat E^a_\mu$ as a result of the spontaneous symmetry breaking:
\begin{equation}
e^a_\mu=<\hat E^a_\mu>\,.
\label{TetradsFermions}
\end{equation}
This order parameter breaks the separate symmetries under orbital and spin transformations, but remains invariant under the combined rotations. On the level of the Lorentz symmetries the symmetry breaking scheme is $L_L \times L_S \rightarrow L_J$. Here $L_L $ is the group of Lorentz transformations in the coordinates space, $L_S $ is the group of Lorentz transformations in the spin space, and 
$L_J$ is the symmetry group of the order parameter, which is invariant under the combined Lorentz transformations $L_J$. 

Similar symmetry breaking mechanism of emergent gravity  is known in condensed matter physics, where the effective gravitational vielbein also emerges as the bilinear fermionic 1-form \cite{Volovik1990}.
This scenario takes place in the $p$-wave spin-triplet superfluid $^3$He-B, where the corresponding relative symmetry breaking \cite{Leggett1973} occurs between the spin and orbital rotations, $SO(3)_L \times SO(3)_S \rightarrow SO(3)_J$. This means that the  symmetry under the relative rotations in spin and orbital spaces is broken, while the properties of $^3$He-B are isotropic.

\section{Dimensionful metric and dimensionless interval}

The metric field is the bilinear combination of the tetrad fields:
 \begin{equation}
g_{\mu\nu}=\eta_{ab}e^a_\mu e^b_\nu \,,
\label{DimMetric}
\end{equation}
and thus in this quantum gravity the metric is the fermionic quartet (in principle the signature can be the dynamical variable $O_{ab}$,\cite{BondarenkoZubkov2022,Bondarenko2022} and $\eta_{ab}$ may also emerge as the vacuum expectation value of the corresponding symmetry breaking phase transition, $\eta_{ab}=<O_{ab}>$.

Ii is important that in this quantum gravity, the fermionic fields $\Psi$ are dimensionless, since they are normalized by the Berezin integral \cite{VladimirovDiakonov2012}. Thus the tetrads in Eq.(\ref{TetradsFermions}) have the dimensions of the inverse time and inverse length, $[e^a_0]=1/[t]$ and
 $[e^a_i]=1/[L]$, while the metric elements in Eq.(\ref{DimMetric}) have dimensions $1/[t]^2$, $1/[L]^2$
 and $1/[t][L]$.
 Due to such dimensions of tetrads and metric, the interval  is dimensionless:
  \begin{equation}
ds^2=g_{\mu\nu}dx^\mu dx^\nu \,\,, \,\,  [s^2]=[1] \,.
\label{DimensionInterval}
\end{equation}
The reason for that is that  the interval is the diffeomorphism invariant, while in this approach to quantum gravity all the
diffeomorphism-invariant quantities are dimensionless \cite{VladimirovDiakonov2012}.

The same takes place for the other  diffeomorphism invariant quantities: the action $S$ (example is in Eq.(\ref{OriginalActions})); scalar curvature $R$; scalar field $\Phi$; the wave function $\psi$; masses $M$;  cosmological constant $\Lambda$; etc.\cite{Volovik2021,Volovik2021cont}. This is valid for the arbitrary dimension of spacetime, and thus is universal, which is one of the most important consequences of the composite tetrads.

Note that the original action (\ref{OriginalActions}) does not contain the Planck constant $\hbar$. One can show that this is the property of any action, if it is written  in the diffeomorphism invariant form, see Section \ref{nohbar}.
 
\section{Action, mass and scalar field are dimensionless}

Let us consider the simplest example of the dimensionless action --  the action describing interaction of a charged point particle with the $U(1)$ gauge field:
 \begin{equation}
S=q \int dx^\mu A_\mu \,.
\label{ChargeParticleAction}
\end{equation}
 As the original action (\ref{OriginalActions}), this action does not depend on the metric field and is described solely in terms of differential forms, now in terms of 
 the 1-form $U(1)$ gauge field $A_\mu$. The $U(1)$ field is the geometric quantity, which comes from the gauging of the global $U(1)$ field.  The field $A_\mu$ comes from the gauging of gradient of the phase field, and thus has dimension of the gradient of phase, with $[A_0]=1/[t]$ and $[A_i]=1/[L]$.
The charge  $q$ here is dimensionless -- it is the integer (or fractional) geometric charge of the fermionic or bosonic field.  As a result the action (\ref{ChargeParticleAction}) is naturally dimensionless, $[S]=1$.

Such action can be extended to the objects of higher dimensions, which interact with the corresponding gauge fields:  $1+1$ strings interacting with 2-form gauge field, $2+1$ branes interacting with the 3-form field and also $3+1$ medium interacting with the 4-form field.

Now let us consider the action describing the classical dynamics of a point particle. This action requires  the metric field, since it is expressed in terms of the  interval:
 \begin{equation}
S=M\int ds \,\,,\,\, ds^2 = -g_{\mu\nu}dx^\mu dx^\nu \,.
\label{particleAction}
\end{equation}
Since both the interval $ds$ and the action $S$ are  dimensionless, from equation (\ref{particleAction})  it  follows that the particle mass  $M$ is also dimensionless, $[M]=[S]=[s]=[1]$.

Let us consider the quadratic terms in the action for the classical scalar field $\Phi$:
\begin{equation}
S=\int d^4 x\,\sqrt{-g} \,\left(  g^{\mu\nu} \nabla_\mu \Phi^*  \nabla_\nu \Phi +M^2|\Phi|^2 \right)\,.
\label{scalar}
\end{equation}
Comparing the gradient and the mass terms, and using the dimension of the metric, one again obtains that the mass $M$ is dimensionless, $[M]=[1]$.
Then since the action $S$ and volume element  $d^4 x\,\sqrt{-g}$ are dimensionless, it follows that the scalar field is also dimensionless,  $[\Phi]^2=[M]=[S]=[1]$.

\section{Schr\"odinger equation in Minkowski spacetime and two Planck constants}

Expanding the Klein-Gordon equation for scalar $\Phi$ in Eq.(\ref{scalar}) over $1/M$ one obtains the non-relativistic Schr\"odinger action.  In Minkowski spacetime, introducing the Schr\"odinger wave function $\psi$:
\begin{equation}
\Phi({\bf r},t) = \frac{1}{\sqrt{M}}\exp\left(i Mt /\sqrt{-g^{00}}\right)\psi({\bf r},t)  \,,
\label{eq:PhiPsi}
\end{equation}
 one obtains the Schr\"odinger-type  action in the form
\begin{eqnarray}
S_{\rm Schr}=\int d^3x dt  \sqrt{-g}\, {\cal L}_{\rm q} \,,
\label{eq:SchroedingerAction}
\\
2{\cal L}_{\rm q}= 
i\sqrt{-g^{00}} \left(\psi \partial_t \psi^*-\psi^* \partial_t \psi\right) +
\nonumber
\\
+\frac{g^{ik}}{M}\nabla_i\psi^* \nabla_k \psi 
+2U|\psi|^2  \,.
\label{eq:SchroedingerEq}
\end{eqnarray}
Here we added the potential term with $U= \sqrt{-g^{00}} qA_0$, where $A_0$ of the electromagnetic gauge field and $q$ is the geometric  charge 
of the scalar field.

Eq.(\ref{eq:SchroedingerEq}) suggests that the metric element $\sqrt{-g^{00}}$ of the Minkowski vacuum plays the role  of the Planck constant $\hbar$. This connection between $g_{00}$ and $\hbar$ was also suggested in Ref. \cite{Volovik2009}, where it was noticed that if $\hbar$  is absorbed into Minkowski metric it does not enter equations written in the covariant form. 
Since in the AD approach to quantum gravity the interval is dimensionless, the Planck constant has dimension of time, $[\hbar]=[t]$. But the term with the space gradients suggests that spatial elements of the Minkowski metric play the roles of another Planck constant, which we denote as slash $\nh$:
 \begin{equation}
-g^{00}_{\rm Mink} \equiv\hbar^2\,\,,\,\, g^{ik}_{\rm Mink}\equiv\nh^2 \delta^{ik}\,.
\label{MinkowskiMetric}
\end{equation}
These Planck constants,  $\hbar$ and $\nh$, enter correspondingly the time derivative and space derivative terms in Schr\"odinger equation:
\begin{equation}
i\hbar \partial_t \psi =-\frac{\nh^2}{2M} \nabla^2\psi + U\psi \,.
\label{SchrodingerEq}
\end{equation}
$\hbar$ and $\nh$ have different dimensions:
\begin{equation}
[\hbar]=[t] \,\,,\,\,  [\nh]=[L] \,,
\label{hdimensions}
\end{equation}
and their ratio $\nh/\hbar$ determines the speed of light $c$ in Minkowski vacuum. Eq.(\ref{hdimensions}) suggests that the Planck constants represent the units of space and time, rather than the units of the phase space.

 All the terms in the Schr\"odinger equation (\ref{SchrodingerEq}) are dimensionless, including the potential energy $U$.  This can be checked for the Coulomb potential for electron with the geometric charge $q=-1$ in the field of nuclear with the geometric charge $q=Z>0$. This potential has the conventional form $U(r)= - Ze^2/r$, where $e$ is the "physical charge", which can be expressed in terms of the fine structure constant.
The fine structure constant is diffeomorphism invariant and thus is dimensionless. That is why from equation
\begin{equation}
\alpha=\frac{e^2}{\nh} \,,
\label{FineStructure}
\end{equation}
it follows that $e^2=\nh \alpha$ has dimension of length, 
\begin{equation}
 [e^2]=[\nh]=[L] \,,
\label{e2}
\end{equation}
and thus the potential $U(r) = -Ze^2/r$ is dimensionless, $[U]=[1]$.
Then one has
\begin{equation}
i\hbar \partial_t \psi ={\cal H}\psi \,\,,\,\, {\cal H}= -\frac{\nh^2}{2M} \nabla^2 - \alpha Z \frac{\nh}{r}  \,.
\label{SchrodingerEq2}
\end{equation}
 The other possible potential terms are also dimensionless. For example, the dipole term $U_{\rm dip}\sim d^2/r^3$ is dimensionless, since $[d^2]=[e^2][L^2]=[L]^3$. The Pauli term for electron
 $U_{\rm P}=  -\frac{\nh^2}{M_e} {\bf B} \cdot \boldsymbol{\sigma}$  is dimensionless, since
 $[\nh]=[L]$, the geometric magnetic field $[B]=1/[L]^2$, and the electron mass $[M_e]=1$.

Note that the Hamiltonian ${\cal H}$, which enters the Schr\"odinger equation, comes from the variation of the dimensionless action over the dimensionless $\psi^*$ and thus is dimensionless, $[{\cal H}]=1$. On the other hand, the Hamiltonian, which comes from the action as $H=dS/dt$, has dimension of frequency, $[H]=1/[t]$. The relation between the energy and frequency will be discussed in Section \ref{EnergyFrequency}.

 Schr\"odinger equation contains two Planck constants with different dimensions, while the other parameters, such as $M$ and $\alpha$, are dimensionless. 

\section{From quantum vacuum to classical physics via symmetry breaking}

From Eq.(\ref{MinkowskiMetric}) it follows that in the Minkowski  spacetime the tetrads have the following values:
\begin{equation}
 e^\mu_a ={\rm diag}(-\hbar,\nh,\nh,\nh) \,\,,\,\,  e_\mu^a ={\rm diag}(-\frac{1}{\hbar}, \frac{1}{\nh}, \frac{1}{\nh}, \frac{1}{\nh})\,,
\label{MinkowskiTetrad}
\end{equation}
Here $e^\mu_a$ are tetrads that are inverse to $e_\mu^a$. They have dimension of length, $[e^i_a]=[L]$, and time $[e^0_a]=[t]$. Their determinant (which is inverse to the determinant $e$ of  $ e_\mu^a$) has dimension of the 4-volume. As a result, the vacuum expectation value of the original action (\ref{OriginalActions}) serves as the number of the elementary 4-volumes.

From Eq.(\ref{TetradsFermions}) it follows that the tetrads $e_\mu^a$ represent the order parameter of the symmetry breaking phase transition with $e_\mu^a=0$ in the symmetric phase. This suggests that in the AD approach to quantum gravity the symmetric phase of the vacuum corresponds to $\hbar=\infty$ and  $\nh=\infty$. It is the pure quantum vacuum, with quantum correlations at the infinitely long distances due to scale invariance. The scale invariance is broken by the finite values of $\hbar$ and $\nh$ in the broken symmetry phases, and this finally gives rise to the classical physics for large masses. The classical physics emerges only in the broken symmetry states. The $\hbar$ expansion in the classical limit \cite{Brodsky2011} 
is opposite to the $1/\hbar$ expansion in the quantum limit.

In this respect the Planck constant $\nh$ has analogy with the coherence length $\xi$ in the second order phase transitions in superconductors and superfluids. The scale of $\xi$ is intermediate between the microscopic length scale $a$, the interatomic distance, and the macroscopic scale $l$ of superfluid hydrodynamics: $a \ll \xi \ll l$. In microscopic physics we use $a/\xi$ as small parameter (quantum limit), while in macroscopic physics the small parameter is $\xi/l$ (classical limit).

The action for the massive Dirac particles is
\begin{equation}
S=\int d^4x\,  e\, (ie^\mu_a \bar\Psi  \gamma^a \nabla_\mu \Psi - M\bar\Psi \Psi)\,.
\label{Fermions}
\end{equation}
The Dirac field is dimensionless, $[\Psi]=[1]$, as well as the 4-volume element,
$[d^3x dt \,e]=[1]$.
In the limit of large wavelength, $\lambda \gg \nh/M$, one obtains the Schr\"odinger equation  for the non-relativistic fermions in Eq.(\ref{SchrodingerEq}), and from that equation -- the classical physics of massive particle at large $M$.

Since the Minkowski metric in Eq.(\ref{MinkowskiMetric}) is quadratic in $1/\hbar$ and $1/\nh$, these Planck parameters may have negative signs, which corresponds to the different signs of the tetrad elements in Minkowski vacuum in Eq.(\ref{MinkowskiTetrad}). In principle, there can be the topological objects related to the symmetry breaking, such as the cosmological domain walls between the Minkowski vacua with positive and negative signs of $1/\hbar$  and/or $1/\nh$.\cite{Volovik2009,Volovik2022} Inside the domain wall the symmetric quantum vacuum with $1/\hbar=1/\nh=0$ is restored, or partially restored if only one of the Planck constants changes sign. Example of such walls can be found in Ref.\cite{Vergeles2022}. 
The same takes place in the cores of the other topological objects: torsion strings\cite{Volovik2022} and topological instantons.\cite{Regge1978,Regge1982} 
Analytic extension of $1/\hbar$ and $1/\nh$ across the Big Bang is also possible, which is similar to the analytic extension of metric in Refs. \cite{Boyle2022,Boyle2022a}.

\section{Energy and frequency}
\label{EnergyFrequency}

Since the action is dimensionless, it may serve as the phase of the wave function in path integral presentation or in the path integrals over the quantum fields. For point particle one has
 \begin{equation}
e^{iS}=e^{iM\int ds} \,.
\label{exponent}
\end{equation}
Let us consider the particle at rest in the Minkowski vacuum:  
 \begin{equation}
e^{iS}=e^{i\int {\cal L}(t) dt}=e^{iMt\sqrt{-g_{00}}} 
 \,.
\label{Stationary}
\end{equation}
This function is the periodic in time with period
  \begin{equation}
{\cal T}=\frac{2\pi}{M\sqrt{-g_{00}}}\,,
\label{period}
\end{equation}
which corresponds to the frequency of oscillations:
  \begin{equation}
\omega=M\sqrt{-g_{00}}\,.
\label{frequency}
\end{equation}
The quantum mechanical relation between energy of stationary particle and frequency, $M=\hbar \omega$, demonstrates again, that the Planck constant $\hbar$ can be considered as the element of Minkowski metric: 
 \begin{equation}
\hbar=\frac{1}{\sqrt{-g_{00}^{\rm Mink}}} = \sqrt{-g^{00}_{\rm Mink}}\,.
\label{hbarfirst}
\end{equation}

Note that the Planck constant was introduced by Planck as a quantum of action. But now, since the action is dimensionless, 
the quantum of action is also dimensionless, $\Delta S= 2\pi$ (or $\Delta S= \pi$ for fermions). Nevertheless, the main property of the Planck constant remains valid: the $\hbar$ enters the relation between the energy and frequency, $M=\hbar \omega$. But now it has dimension of time, $[\hbar]=1/[t]$.

\section{de Sitter spacetime and Planck constants}

The same parameters $\hbar$ and $\nh$ exist for any $D+1$ Minkowski spacetime.
However, these parameters are not diffeomorphism invariant. Being the element of the Minkowski metric they are invariant only space rotations. As a result the Planck constants are not dimensionless. Then according to Weinberg criterion \cite{WeinbergCriterium}  they cannot be the fundamental constants  (see also Refs. \cite{Trialogue,OkunCube,OkunGamov,Gamov} on fundamental constants).

Let us consider the possible variation of the Planck constants on example of the  de Sitter (dS) spacetime. The dS spacetime can be obtained from the 4+1 Minkowski spacetime:
\begin{equation}
-\frac{1}{\hbar^2}dt^2   +\frac{1}{\nh^2}\sum_1^4 X^iX^i =\alpha^2\,.
\label{dS}
\end{equation}
It contains one more parameter, the dimensionless constant $\alpha$, the radius of the 4+1 sphere.
The corresponding Hubble parameter has dimension of frequency:
\begin{equation}
H=\frac{1}{\hbar \alpha}\,\,, \,\, [H]=\frac{1}{[t]}\,.
\label{dS_H}
\end{equation}
 In the Paineve-Gullstrand form, the interval in dS spacetime contains three parameters: $\hbar$, $\nh$ and $H$:
\begin{equation}
ds^2 =-\frac{1}{\hbar^2}dt^2  +\frac{1}{\nh^2}\left( (dr -Hrdt)^2 + r^2 d\Omega^2 \right)\,.
\label{dSds}
\end{equation}
At $r=0$ the metric is Minkowski. However, since the Planck constants are not fundamental, it is not excluded that in the dS Universe they may deviate from their values in Minkowski vacuum and depend on $H$. The phonon analog of the metric emerging in liquids suggests the following corrections to the Planck constants:\cite{Volovik2023c}
\begin{equation}
\frac{\Delta \nh}{\nh} \sim \frac{\Delta \hbar}{\hbar} \sim\hbar^2H^2 =\frac{\nh^2}{r_c^2} \ll1\,.
\label{Variation}
\end{equation}
Here $r_c$ is the radius of the cosmological horizon.

Note that the main cosmological constant problem is not affected by this dependence of $\hbar$. In the $q$-theory of the quantum vacuum\cite{KlinkhamerVolovik2008b} the vacuum energy is self-tuned to zero in the full equilibrium, and this does not depend on the value of $\hbar$. 

\section{Black hole and Planck constants}

Let us consider the possible variation of the Planck constants coming from the black hole.
The black hole metric in the Paineve-Gullstrand form is
 \begin{equation}
ds^2 =-\frac{1}{\hbar^2}dt^2  +\frac{1}{\nh^2}\left( (dr -vdt)^2 + r^2 d\Omega^2 \right)\,,
\label{BHds}
\end{equation}
where $v(r)$ is the corresponding shift function.
The radius of the black hole event horizon is
 \begin{equation}
r_h=2 MG\,,
\label{BHv}
\end{equation}
where $M$ is the dimensionless mass of the black hole and  $G$ is the Newton constant, which has dimension of length, see Section \ref{LengthLength}. 
So the metric also contains 3 parameters: $\hbar$, $\nh$ and  the parameter $MG$. The metric becomes Minkowski at $r\rightarrow \infty$. But near the horizon it may deviate from the vacuum values.
The comparison with the Eq.(\ref{Variation}), where the corrections to the Planck constants are inverse proportional to the square of the event horizon radius, suggests the following corrections:
\begin{equation}
\frac{\Delta \nh}{\nh} \sim \frac{\Delta \hbar}{\hbar} \sim \frac{\nh^2}{r_h^2} \ll1\,.
\label{Variation2}
\end{equation}

\section{Planck constants and Tolman law}

In the dS spacetime, the probability of Hawking radiation of particle with mass $M$ detected by observer at $r=0$ is determined by parameters $\hbar$ and $H$:
\begin{eqnarray}
w \propto \exp\left(- \frac{2\pi  M}{H \sqrt{-g^{00}_{\rm Mink}}} \right)=\exp\left(- \frac{2\pi  M}{\hbar H} \right)=
\label{HawkingdS1}
\\
=\exp\left(- \frac{M}{T_H} \right)\,.
\label{HawkingdS2}
\end{eqnarray}
Here $T_H$ is the Gibbons-Hawking temperature measured at $r=0$, where the metric is Minkowski:
\begin{equation}
T_H=T(r=0)=\frac{ \sqrt{-g^{00}_{\rm Mink}} H}{2\pi}=\frac{\hbar H}{2\pi}\,.
\label{HawkingTdS}
\end{equation}
This temperature is dimensionless due to the time dimension of the Planck constant:  $[T_H]=[\hbar][H]=[t]/[t]=[1]$.

On the other hand, the parameter $H/2\pi$ plays the role of Tolman temperature, which enters the Tolman law:
\begin{equation}
T(r)=\frac{T_{\rm Tolman}}{ \sqrt{-g_{00}(r)} } \,\,, \,\,  T_{\rm Tolman}=\frac{H}{2\pi}\,,
\label{HawkingTolman}
\end{equation}
has dimension of inverse time, $[T_{\rm Tolman}]=[H]=1/[t]$, see also Refs \cite{Volovik2021,Volovik2021cont}.

The Hawking temperature of black hole, which is measured at the asymptotic Minkowski vacuum, is
 \begin{equation}
T_H=T(r=\infty)=\frac{\nh}{4\pi  r_h}\,,
\label{TBH}
\end{equation}
where $r_h$ is the position of the black hole horizon.
As the Gibbons-Hawking temperature in Eq.(\ref{HawkingTdS}), the Hawking temperature (\ref{TBH}) is also dimensionless, now  due to  the length dimension of the second Planck constant, $[T_H] =[\nh]/[r_h]= [L]/[L]=[1]$.

It is not excluded that the Tolman temperature is the parameter of the equilibrium system, which may influence the Planck constants. The equations (\ref{Variation}) and (\ref{Variation2}) suggest the following corrections to the Planck constants:
\begin{equation}
\frac{\Delta \nh}{\nh} \sim \frac{\Delta \hbar}{\hbar} \sim \hbar^2 T_{\rm Tolman}^2 \ll 1\,.
\label{Variation3}
\end{equation}

\section{Length dimension of Newton constant and Planck length}
\label{LengthLength}

In the composite fermion gravity, the gravitational potential $U(r)=-GM_1M_2/r$ is dimensionless and contains masses $M_1$ and $M_2$, which are also dimensionless. As a result the Newton constant has the dimension of length, $[G]=[L]$. This suggests, that $G$ is not diffeomorphism invariant, and thus cannot be the fundamental constant. That is why in the gravitational action it must be compensated by  $\nh$, which also has dimension of length, $[G]=[\nh]=[L]$: 
\begin{eqnarray}
\hspace*{-10mm}
S = \frac{1}{16\pi} \frac{\nh}{G} \int d^4x\,\sqrt{-g} \,R  \,.
\label{EinsteinAction4D}
\end{eqnarray}
Since the scalar curvature $R$ is dimensionless, the Einstein–Hilbert action (\ref{EinsteinAction4D}) is dimensionless.
It can be written via  diffeomorphism invariant quantities if we introduce  the Planck mass $M_{\rm P}=\sqrt{\nh/G}$, which is dimensionless as all the masses $M$ in the composite tetrad approach, $[M_{\rm P}]^2=[\nh]/[G]=[L] [L]^{-1}=[1]$: 
\begin{eqnarray}
\hspace*{-10mm}
S = \frac{M_{\rm P}^2}{16\pi}  \int d^4x\,\sqrt{-g} \,R  \,.
\label{EinsteinActionMP}
\end{eqnarray}

 The Planck length scale has the conventional form  $l^2_{\rm P}= \nh G$, with $[l_{\rm P}]^2= [\nh] [G]=[L] [L]=[L]^2$.
The slash Planck constant $\nh$ has the same dimension as the Planck length, $[\nh]=[l_{\rm P}]=[L]$. Whether this "Planck constant length" is related to the "Planck length scale", is an open question \cite{Carlip2022}.  This question was considered on the example of the acoustic gravity, where the analog of the trans-Planckian phyics -- atomic physics  -- is known \cite{Volovik2023c}. It was demonstrated  that the acoustic analog of $\nh$ is on the order of the interatomic distance. This suggests that in quantum gravity the Planck constant $\nh$ is on the order of Planck length $l_{\rm P}$, i.e. the Planck mass is on the order of unity, $M_{\rm P} =\sqrt{\nh/G} \sim 1$.

\section{No $\hbar$ and  $\nh$ in diffeomorphism invariant equations}
\label{nohbar}

Let us consider the diffeomorphism invariant equations on example of the statistical entropy, which is dimensionless in any units. The Gibbons-Hawking entropy of the de Sitter cosmological horizon is
  \begin{equation}
S_H=\frac{M_{\rm P}^2}{4\pi T_H^2}\,,
\label{dSentropy}
\end{equation}
where $T_H$ is the Gibbons-Hawking temperature in Eq.(\ref{HawkingTdS}).
The Bekenstein-Hawking entropy of the black hole is
  \begin{equation}
S_H=\frac{4\pi M^2}{M_{\rm P}^2}= \frac{M}{2T_H}\,,
\label{BHentropy}
\end{equation}
 where $M$ is the black hole mass and $T_H$ is the Hawking temperature of black hole radiation.
 
 All quantities in Eqs. (\ref{dSentropy}) and (\ref{BHentropy}) are dimensionless, $[S_H]=[M_{\rm P}]=[T_H]=[M]=[1]$. Both equations do not contain Planck constants. This demonstrates the general property of  diffeomorphism invariant equations: they do not contain  $\hbar$ and  $\nh$, because the Planck
 constants are not diffeomorphism invariant and have dimensions of time and length correspondingly.

\section{Conclusion}

The important consequence of the composite tetrads approach to quantum gravity is the "dimensionless physics": all the diffeomorphism invariant quantities are dimensionless for any dimension of spacetime. These include the action $S$, interval $s$, cosmological constant $\Lambda$, Hawking temperature $T_H$, scalar curvature $R$, scalar field $\Phi$, 
Planck mass $M_{\rm P}$, masses $M$ of particles and fields, etc.  

Another consequence of this approach to quantum gravity, is that there are two Planck constants, bar $\hbar$ and slash $\nh$, which are the elements of the Minkowski metric.  As the elements of the Minkowski metric, $\hbar$ and $\nh$ are invariant only under $SO(3)$ space rotations, and thus they are not diffeomorphism invariant.  As a result the Planck constants are not dimensionless, with bar $\hbar$ having dimension of time, $[\hbar]=[t]$, and slash $\nh$ having dimension of length $[\nh]=[L]$, and they do not enter the diffeomorphism invariant equations.

Since the Planck constants are not dimensionless, then according to Weinberg criterion they cannot be the fundamental constants, and thus may vary with space and time. The possible corrections to the Planck constants in the de Sitter Universe and near the event horizon of black hole are in Eqs.(\ref{Variation}) and
(\ref{Variation2}).

 According to Vladimirov and Diakonov, \cite{VladimirovDiakonov2012} "the unconventional dimensions of the fields ... are natural and adequate for a microscopic theory of quantum gravity".
The similar "dimensionless physics" appears also in several other approaches to quantum gravity. It appears in particular in the $BF$- theories, where the composite metric is formed by the triplet of the 2-form fields  (Sch\"onberg-Urbantke metric)\cite{Schonberg1971,Urbantke1984,Jacobson1991,Obukhov1996,HehlObukhov2003,Friedel2012}.  It appears also in the model of the superplastic vacuum,\cite{KlinkhamerVolovik2019} which is described in terms of the so-called elasticity tetrads\cite{Dzyaloshinskii1980,NissinenVolovik2019,Nissinen2020,Nissinen2020a,Nissinen2020b,Burkov2021,Burkov2023}, and in acoustic gravity \cite{Volovik2023c}. All this suggests that the physics with two Planck constants, bar  $\hbar$ and slash $\nh$,  can be reasonable.

\acknowledgments
 I thank Y.N. Obukhov for discussion and criticism.  This work has been supported by Academy of Finland (grant 332964).

\end{document}